\documentclass[12pt]{article}
\usepackage{geometry}                
\usepackage{graphicx}
\usepackage{amssymb}
\usepackage{epstopdf}
\usepackage{color}

\usepackage{amsmath}
\usepackage{graphics,graphicx,epsfig}
\usepackage{amssymb}
\usepackage{sidecap}
\sidecaptionvpos{figure}{c}
\usepackage{appendix}
\usepackage{cite}
\usepackage{multicol}
\usepackage{array}
\newcolumntype{o}{@{}>{{}}c<{{}}@{}}

\titlepage
\title{Lovelock branes}

\usepackage{geometry}                
\usepackage{graphicx}
\usepackage{amssymb}
\usepackage{epstopdf}
\usepackage{color}

\usepackage{amsmath}
\usepackage{graphics,graphicx,epsfig}
\usepackage{amssymb}
\usepackage{sidecap}
\sidecaptionvpos{figure}{c}
\usepackage{appendix}

\def\be{\begin{equation}}
\def\ee{\end{equation}}

\begin{document}

\begin{titlepage}
\vfill
\vskip 3.0cm
\begin{flushright}
ACFI-T17-14
\end{flushright}
\vskip 1.0in

\begin{center}
{\Large\bf Lovelock Branes}
\vskip 10.mm

{\bf  David Kastor${}^{a}$, Sourya Ray${}^{b}$, Jennie Traschen${}^{a}$} 

\vskip 0.4cm
${}^a$ Amherst Center for Fundamental Interactions, Department of Physics\\ University of Massachusetts, Amherst, MA 01003\\

\vskip 0.1in ${}^b$ Instituto de Ciencias F\'{\i}sicas y Matem\'{a}ticas, Universidad Austral de
Chile, Valdivia, Chile\\
\vskip 0.1 in Email: \texttt{kastor@physics.umass.edu, ray@uach.cl, traschen@physics.umass.edu}
\vspace{6pt}

\end{center}
\vskip 0.2 in

\begin{center}
{\bf Abstract}

\end{center}

\begin{quote}

We study the problem of finding brane-like solutions to Lovelock gravity, adopting a general approach
to establish conditions that a lower dimensional base metric 
 must satisfy in order that a solution to a given Lovelock theory can be constructed in one higher dimension. We find that 
for Lovelock theories with generic values of the coupling constants, the Lovelock tensors (higher curvature generalizations of the Einstein tensor) of the base metric must all be proportional to the metric. Hence, allowed base metrics form a 
subclass of Einstein metrics. This subclass includes so-called `universal metrics', which have been previously investigated as solutions to quantum-corrected field equations.
 For specially tuned values of the Lovelock couplings, we find that the Lovelock tensors of the base metric
need to satisfy fewer constraints. For example, for Lovelock theories with a unique vacuum there is
 only a single such constraint, a case previously identified in the literature, and brane solutions can be straightforwardly constructed.

   \vfill
\vskip 2.mm
\end{quote}
\hfill
\end{titlepage}

\section{Introduction}

Vacuum general relativity has the property that if a metric $\hat g_{\mu\nu}(x^\rho)$, which we will call the base metric, solves the field equations in $N$ dimensions, then the metric
\be\label{addone}
ds^2 = dz^2 +\hat g_{\mu\nu}dx^\mu dx^\nu
\ee
is a solution to the field equations in $D=N+1$ dimensions.  This allows for a simple construction of extended black brane solutions based on lower dimensional black holes.  Although this construction seems trivial, the new spacetime can exhibit surprising new properties.  For example, the $D=5$ black string obtained by adding a flat direction to the 4-dimensional Schwarzschild spacetime turns out to be unstable, despite the stability of the black hole in the base metric \cite{Gregory:1993vy}. 

Lovelock gravity \cite{Lovelock:1971yv} is an interesting higher curvature generalization of general relativity that has been studied extensively in a wide variety of physical contexts, {\it e.g.} brane world models and various AdS/CFT applications\footnote{The literature on these subjects is extensive.  See {\it e.g.} \cite{Charmousis:2002rc,Brigante:2007nu,Barclay:2010up} for a few examples of applications.}.  Static, spherical black hole solutions of Lovelock theories have been known for some time, starting with references \cite{Boulware:1985wk,Wheeler:1985nh,Wheeler:1985qd} (see \cite{Charmousis:2008kc,Garraffo:2008hu} for reviews).  One would also like to study extended black branes in Lovelock theories.  However, these have proven difficult to find analytically\footnote{Numerical solutions for static Lovelock black branes were found in \cite{Kobayashi:2004hq}.}.  In particular, it is known that the construction (\ref{addone}) does not work in generic Lovelock theories \cite{Barcelo:2002wz}.  

This latter property should not be surprising.   If one simply adds a `lower curvature' interaction to vacuum general relativity,  a non-vanishing cosmological constant $\Lambda$ so that the gravitational action is
\be\label{einstein}
S= {1\over 16\pi G}\int d^Dx(R-2\Lambda)
\ee
then the simple construction (\ref{addone}) no longer holds.  In this case, however, there is an only slightly more complicated construction that does work. Assume for definiteness that  $\Lambda<0$ and write it in terms of a length scale $l$ according to
\be\label{cosmo}
\Lambda =-{(D-1)(D-2)\over 2l^2}
\ee
Consider ``warped" metrics of the form
\begin{align}\label{ansatz}
ds^2=\dfrac{dz^2}{g(z)}+{z^2\over l^2}\hat{g}_{\mu\nu}dx^{\mu}dx^{\nu}
\end{align}
It is well known that if the $N$-dimensional Lorentzian base metric $\hat g_{\mu\nu}(x^\rho)$ satisfies the vacuum Einstein equations, then the $D=N+1$ dimensional metric (\ref{ansatz}) with $g(z) = z^2/l^2$ will solve the Einstein equations with cosmological constant $\Lambda$.  For example, if the base metric is Minkowski spacetime, then the extended metric is AdS spacetime, while if the base metric is Schwarzschild, this construction gives the AdS black string \cite{Chamblin:1999by}.  

More generally,  a solution  (\ref{ansatz}) can be found whenever $\hat g_{\mu\nu}$ is an Einstein metric, so that it is a solution to the theory
%
\be
S= {1\over 16\pi G}\int d^Nx(\hat R-2\Lambda^\prime)
\ee
where $\hat R$ is the Ricci scalar for the base metric and $\Lambda^\prime$ is an arbitrary constant.
Assuming, again for definiteness, that $\Lambda^\prime\le 0$ and corresponds to a curvature length scale $l^\prime$ defined via the analogue of (\ref{cosmo}),  then the $D$-dimensional equations of motion will be satisfied for
\be\label{simpleg}
g(z) = {z^2\over l^2} - {l^2\over l^{\prime 2}}
\ee
A similar expression for $g(z)$ holds for $\Lambda^\prime>0$.  

In this paper we will ask an analogous question for solutions to Lovelock gravity.  Namely, for what class of $N$-dimensional base metrics $\hat g_{\mu\nu}$ does the warped ansatz (\ref{ansatz}) yield a solution to the $D=N+1$ dimensional Lovelock equations of motion.  In the generic Lovelock case, meaning for general choices of the coupling constants, we find that warped solutions can be found if the base metric $\hat g_{\mu\nu}$ is what we will refer to as a ``Lovelock" metric, a property which we will define in close analogy with Einstein metrics.  For special tunings of the coupling constants, we find that  additional, less constrained, possibilities for the base metric also exist\footnote{Previous work \cite{Kastor:2006vw,Giribet:2006ec} has considered distinct, but closely related, questions. We indicate below how our present results mesh with these earlier studies.  See also  \cite{Barcelo:2002wz} for other related results.}.  

The paper proceeds as follows.  In section (\ref{basic}) we describe the basics of Lovelock gravity theories and derive the equations of motion for the metric ansatz (\ref{ansatz}).  In section (\ref{einstein}) we warm up by examining these equations for Einstein gravity, regarded as a special case of Lovelock gravity, and show how the results quoted above emerge in this formalism.  In section (\ref{general}) we explore some general properties of the Lovelock equations of motion for the ansatz (\ref{ansatz}) that will be helpful in the analysis.  In section (\ref{lovelockspaces}) we define higher order Lovelock spaces in analogy with Einstein spaces, which are $1$st order Lovelock spaces, and show that solutions of the form (\ref{ansatz}) can be found for any Lovelock theory, with generic values of the coupling constants, if the base metric is a Lovelock space.  In section (\ref{polynomialsection}) we explore a different class of solutions, which we call polynomial solutions, which arise when the Lovelock coupling constants take special non-generic values, and which have weaker constraints on the base metric\footnote{We expect that these two classes of solutions, the polynomial solutions and those based on Lovelock spaces, exhaust the entire class of solutions of the form (\ref{ansatz}), but have not so far been able to demonstrate this.}.  In section (\ref{example}), as an example, we work out the different conditions that are imposed on the base metric for special sets of couplings in third order Lovelock theory.  Finally, we offer some concluding remarks in section (\ref{conclusions}).


\section{Basic setup and calculations}\label{basic}

Lovelock gravity \cite{Lovelock:1971yv} is the most general theory constructed from powers of the curvature tensor and its contractions having field equations that, like the Einstein equations, are second order in derivatives of the metric.   The action is given by
\be\label{lovelock}
S = {1\over 16\pi \tilde G}\int d^Dx\sqrt{-g}\,\left( \sum_{k=0}^{\bar k}c_k\mathcal{L}^{(k )}\right)
\ee
where the individual Lovelock interaction terms are 
\be\label{scalars}
\mathcal{L}^{(k)}=\dfrac{1}{2^k}\delta_{a_1b_1\cdots a_kb_k}^{c_1d_1\cdots c_kd_k}R_{c_1d_1}^{\ \ \ \ a_1b_1}\cdots R_{c_kd_k}^{\ \ \ \ a_kb_k}
\ee
the delta symbol is a totally antisymmetrized product of Kronecker delta functions, normalized such that its non-zero entries take values $\pm 1$, and the $c_k$'s are dimensionful coupling constants.
The equations of motion for Lovelock gravity have the form
\be\label{eom}
\mathcal{G}^a_c=\sum_{k=0}^{\bar{k}}c_k\mathcal{G}^{(k)a}{}_{c}=0
\ee
where the individual Lovelock tensors are given by
\be\label{elike}
\mathcal{G}^{(k)a}{}_{c}=-\dfrac{1}{2^{k+1}}\delta^{aa_1b_1\cdots a_kb_k}_{cc_1d_1\cdots c_kd_k}R^{\ \ \ \ c_1d_1}_{a_1b_1}\cdots R^{\ \ \ \ c_kd_k}_{a_kb_k}.
\ee
From the antisymmetrization, we see that the $k$th order contribution to the equations of motion is nontrivial only if $D\ge 2k+1$.  This imposes an effective upper limit of $\bar k \le [(D-1)/2]$ to the number of interaction terms in the Lovelock action (\ref{lovelock}).  We note for use below that the trace of the Lovelock tensors is given by
\be
\mathcal{G}^{(k)a}{}_{a}=-{D-2k\over 2}\mathcal{L}^{(k)}
\ee
The first two terms in the Lovelock action (\ref{lovelock}) are the basic interactions of Einstein gravity, with
\be\label{basics}
\mathcal{L}^{(0)}=1,\qquad \mathcal{L}^{(1)}=R,\qquad \mathcal{G}^{(0)a}{}_{c}=-{1\over 2}\delta^a_c,\qquad \mathcal{G}^{(1)a}{}_{c}=G^a{}_c
\ee
Comparing the constants in the Lovelock action with those in the Einstein action (\ref{einstein}), we see that $G=\tilde G/c_1$ and $\Lambda = -c_0/2c_1$.  Because of the upper bound on $\bar k$, Einstein gravity with a cosmological constant is the most general Lovelock theory in four dimensions.  The second order `Gauss-Bonnet' term $\mathcal{L}^{(2)}$ becomes relevant in $D=5$, while the third order term $\mathcal{L}^{(3)}$ first arises in $D=7$.

It is important to note that Lovelock gravity with maximal order $\bar k$ can have up to $\bar k$ locally distinct constant curvature vacua.  The constant curvature form for the Riemann tensor may be written as
\be
R_{ab}{}^{cd} = \lambda \delta_{ab}^{cd}
\ee
where $\lambda$ is a constant.  After plugging this into (\ref{elike}) and carrying out the necessary contractions\footnote{The key tool for this is the identity $\delta^{a_1\dots a_n}_{b_1\dots b_n}\delta_{a_{n-1}a_n}^{b_{n-1}b_n} =2(D-(n-2))(D-(n-1))\delta^{a_1\dots a_{n-2}}_{b_1\dots b_{n-2}}$}, one finds that the equations of motion reduce to a $\bar k$th order polynomial equation for the vacuum curvature $\lambda$
\be\label{constant}
f(\lambda)=\sum_{k=0}^{\bar k} a_k\lambda^k = 0
\ee
where the constants $a_k$ are related to the Lovelock couplings $c_k$ via combinatoric factors according to
\begin{align}\label{rescaled}
a_k=\dfrac{(D-1)!}{(D-2k-1)!}c_k
\end{align}
Only real roots of (\ref{constant}) yield constant curvature vacua.  For $\bar k$ even, there is a region of coupling space with no such vacua.  An interesting special case is when all the roots of the polynomial $f(\lambda)$ coincide, so that
\be
f(\lambda) = c (\lambda-\bar\lambda_0)^{\bar k}
\ee
where $c$ is a constant and $\lambda_0$ is the vacuum curvature.  Comparing with (\ref{constant}) we see that in this unique vacuum case, the constants $a_k$ are given by
\be\label{unique}
a_k= c\binom{\bar k}{k}(-\lambda_0)^{\bar k-k}
\ee
Black hole solutions in unique vacuum Lovelock theories were studied in \cite{Crisostomo:2000bb}.

We now consider a metric of the form (\ref{ansatz}), where for the present $l$ is regarded as an arbitrary length scale\footnote{In the following analysis we shall leave the signature of both the extra dimension and the lower dimensional space arbitrary. Hence, our analysis will be applicable for constructing both brane spacetimes, as well as cosmological ones when the added dimension is timelike.}. 
The non-trivial Riemann curvature components are then given by
\begin{align}
R_{\mu\nu}^{\ \ \lambda\rho}=\dfrac{l^2}{z^2}\left(\hat{R}_{\mu\nu}^{\ \ \lambda\rho}-{g(z)\over l^2}\delta_{\mu\nu}^{\lambda\rho}\right),\qquad 
R_{\mu z}^{\lambda z}=-\dfrac{g'(z)}{2z}\delta_{\mu}^{\lambda}
\end{align}
Plugging these into (\ref{eom}) then yields the field equations for the ansatz (\ref{ansatz})
\begin{align}\label{redfeqns1}
0&=\mathcal{G}^z_z=-\dfrac{l^{D-1}}{2(D-1)!z^{D-1}}\sum\limits_{n=0}^{\bar{k}}\left\{(D-2n-1)!\hat{\mathcal{L}}^{(n)}\right\}\left\{({z\over l})^{D-2n-1}A_n(U)\right\}\\
0&=\mathcal{G}^{\alpha}_{\beta}=\dfrac{l^{D-2}}{(D-1)!z^{D-2}}\sum\limits_{n=0}^{\bar{k}}\left\{(D-2n-2)!\hat{\mathcal{G}}^{(n)\alpha}_{\ \ \ \beta}\right\}\left\{ l\dfrac{d\ }{dz}\left( ({z\over l})^{D-2n-1}A_n(U)\right)\right\}
\label{redfeqns2}
\end{align}
where $\hat{\mathcal{L}}^{(n)}$ and $\hat{\mathcal{G}}^{(n)\alpha}_{\ \ \ \beta}$ are the $n$th order scalar Lovelock interaction terms (\ref{scalars}) and Lovelock tensors (\ref{elike}) evaluated on the base metric $\hat g_{\mu\nu}$, and the functions $A_n(U)$ are polynomials of order $\bar{k}-n$ given by
\begin{align}\label{anu}
A_n(U)=\sum\limits_{k=n}^{\bar{k}}a_k \binom{k}{n}U^{k-n}
=\sum\limits_{p=0}^{\bar{k}-n}a_{n+p}\binom{n+p}{n}U^p
\end{align}
where 
$U=-g(z)/z^2$.  The polynomials $A_n(U)$ are known as Appell polynomials \cite{Roman:1984} and satisfy the recurrence relations $A_n'(U)=(n+1)A_{n+1}(U)$,
so that 
\be\label{highern}
A_n(U)={1\over n!}A_0^{(n)}(U)
\ee
where the superscript on the right hand side denotes the number of derivatives with respect to the variable $U$.

\section{Warm up: the case of Einstein gravity}\label{einsteincase}

In order to orient ourselves, first consider the case of maximal Lovelock order $\bar k=1$, which is simply Einstein gravity with a cosmological constant.  As noted in the introduction, the base metric $\hat g_{\mu\nu}$ in (\ref{ansatz}) can, in this case, be any Einstein metric.  We would like to see how this emerges from equations of motion (\ref{redfeqns1}) and (\ref{redfeqns2}), which reduce respectively in this case to
\begin{align}
-{1\over 2}&\left(a_0+a_1U+{a_1 l^2\over (D-1)(D-2)z^2}\hat R\right)=0\\
-{1\over 2}&\left(a_0+a_1U+{a_1z\over (D-1)}{d\over dz}U\right)\delta^\mu_\nu +{a_1l^2\over (D-1)(D-2)z^2}\hat G^\mu_\nu =0
\end{align}
The first equation can be solved for $U(z)$ giving
\be
U(z) = -{a_0\over a_1}-{l^2\over (D-1)(D-2)z^2}\hat R
\ee
Since the left hand side is assumed to be a function of $z$ only, this implies that the scalar curvature $\hat R$ of the base metric must be constant.  After plugging in this result for $U(z)$ the second equation reduces to 
\be
\hat R^\mu{}_\nu = {1\over D-1}\hat R\,\delta^\mu_\nu
\ee
which implies that the base metric $\hat g_{\mu\nu}$ is an Einstein metric.

\section{General considerations}\label{general}

In order to understand solutions to the field equations for $\bar k>1$, it is useful to step back and look at their general form. We see that both equation (\ref{redfeqns1}) and each tensor component of equation (\ref{redfeqns2}) have the general form 
\begin{align}
\sum\limits_{n=0}^{\bar{k}}u_n(y)v_n(z)=0.
\end{align}
where $y$ collectively represents the coordinates $x^{\mu}$ on the base space. This implies that each of the sets of functions $\{u_n(y)\}$ and $\{v_n(z)\}$ is a linearly dependent set. To see this we take derivatives of the above equation with respect to one of the two variables up to order $\bar{k}$, thereby obtaining a linear system of equations. For example, taking derivatives with respect to $z$, we obtain the following system of linear equations in $\{u_n(y)\}$
\begin{equation}\begin{array}{rororororro}
&u_0(y)v_0(z)           &+& u_1(y)v_1(z)&+&\cdots &+&u_{\bar{k}}(y)v_{\bar{k}}(z)&=&0\\
&u_0(y)v_0^{(1)}(z)     &+& u_1(y)v_1^{(1)}(z)&+&\cdots &+&u_{\bar{k}}(y)v_{\bar{k}}^{(1)}(z)&=&0\\
&\vdots                  && \vdots && &&\vdots &=&\vdots\\
&u_0(y)v_0^{(\bar{k})}(z)&+& u_1(y)v_1^{(\bar{k})}(z)&+&\cdots &+&u_{\bar{k}}(y)v_{\bar{k}}^{(\bar{k})}(z)&=&0
\end{array}\end{equation}
%
This system has a non-trivial solution provided the determinant
\begin{align}
\begin{vmatrix}
v_0(z)  &  v_1(z)  & \cdots &  v_{\bar{k}}(z)\\
v_0^{(1)}(z)  &  v_1^{(1)}(z)  & \cdots & v_{\bar{k}}^{(1)}(z)\\
\vdots     &   \vdots     & \vdots &   \vdots\\
v_0^{(\bar{k})}(z)  &  v_1^{(\bar{k})}(z)  & \cdots & v_{\bar{k}}^{(\bar{k})}(z)
\end{vmatrix}
\end{align}
vanishes. This determinant is simply the Wronskian and its vanishing implies that the functions $\{v_n(z)\}$ constitute a linearly dependent set. Similarly, the functions $\{u_n(y)\}$ also form a linearly dependent set.  A key result  \cite{Aczel:2006fe,Suto:1914} is that the ranks of the Wronskian matrices for the two sets of functions must sum to $\bar{k}+1$. 

Applying this argument, for example, to equation (\ref{redfeqns1}), we can infer that there exists at least one set of constants $\alpha_n$, not all zero, such that
\begin{align}\label{genwheelereqn}
\sum\limits_{n=0}^{\bar{k}}\alpha_n\dfrac{l^{2n}}{z^{2n}}A_n(U)=0.
\end{align}
Further, if the rank of the Wronskian matrix for the set $\{ {l^{2n}\over z^{2n}}A_n(U)\}$ is equal to $m$, then there will be $\bar{k}+1-m$ linearly independent such relations among them.

In the next section, we will consider the case that the functions on the base space in (\ref{redfeqns1}) and (\ref{redfeqns2}) have $\bar k$ linearly independent relations amongst them.  This gives rise to a condition on the base metric that will yield solutions of the form (\ref{ansatz}) for any Lovelock theory.  In sections (\ref{polynomialsection}) and (\ref{example}), we explore cases where the functions on the base space are less constrained.  However, such solutions are limited to special classes of Lovelock theories.

\section{Generic class of solutions: Lovelock spaces}\label{lovelockspaces}

For generic values of the Lovelock coupling constants, we find a class of base metrics $\hat g_{\mu\nu}$ such that solutions of the form (\ref{ansatz}) exist.  We will refer to this class of metrics as Lovelock spaces, because as we will see, they satisfy a more stringent analogue of the condition for Einstein spaces.  These conditions arise if we assume that the Wronskian matrices for the sets of functions $\{(D-2n-1)!\hat{\mathcal{L}}^{(n)}\}$ and 
$\{ {l^{2n}\over z^{2n}}A_n(U)\}$ in equation (\ref{redfeqns1}) have ranks $1$ and $\bar k$ respectively.  This assumption implies that the functions $\hat{\mathcal{L}}^{(n)}$ are all proportional to one another.  Moreover because $\hat{\mathcal{L}}^{(0)}=1$, it implies that all the higher order 
$\hat{\mathcal{L}}^{(n)}$ must also be constant.  If we normalize these constants according to 
%
%
%
\begin{align}\label{sclconstraint}
\hat{\mathcal{L}}^{(n)}=\dfrac{(D-1)!}{(D-2n-1)!}\alpha_n,
\end{align}
where $\alpha_0$ is chosen to be $1$, then equation (\ref{redfeqns1}) reduces to (\ref{genwheelereqn}), which is a polynomial equation for the function $U(z)$.

Proceeding in the same fashion we can solve the remaining field equations (\ref{redfeqns2}) corresponding to the components on the $(D-1)$-dimensional space. First of all, there must exist a unique set of constants $\beta_n$, not all zero, such that
\begin{align}
\sum\limits_{n=0}^{\bar{k}} \beta_nl\dfrac{d\ }{dz}\left(({z\over l})^{D-2n-1}A_n(U)\right)=0
\end{align}
Secondly, all the Lovelock tensors $\hat{\mathcal{G}}^{(n)\alpha}_{\ \ \ \beta}$, intrinsic to the $(D-1)$-dimensional space then must be constant multiples of the metric given by
\begin{align}
\hat{\mathcal{G}}^{(n)\alpha}_{\ \ \ \beta}=-\dfrac{1}{2}\dfrac{(D-2)!}{(D-2n-2)!}\beta_n \delta^{\alpha}_{\beta}
\end{align}
where the overall proportionality constant is chosen to fix $\beta_0=1$. Taking the trace of the above relation and using the identity $\hat{\mathcal{G}}^{(n)\alpha}_{\ \ \ \alpha}=-(D-2n-1)!\hat{\mathcal{L}}^{(n)}/2$, we conclude $\alpha_n=\beta_n$. This along with the solution to equation (\ref{genwheelereqn}) solves all the field equations.

Therefore, in order to construct a $D=N+1$ dimensional solution in Lovelock theory of order $\bar{k}$ with generic coupling constants, by stacking $N$-dimensional spaces along an extra dimension, all the (non-trivial) Lovelock tensors instrinsic to the $N$-dimensional slices must be set to multiples of the metric i.e.,
\begin{align}\label{tensconstraint}
\hat{\mathcal{G}}^{(n)\alpha}_{\ \ \ \beta}=-\dfrac{1}{2}\dfrac{(D-2)!}{(D-2n-2)!}\alpha_n\delta^{\alpha}_{\beta}.
\end{align}
We shall call such spaces, for which the Lovelock tensors of all orders upto $\bar{k}$ are constant multiples of the metric, {\it Lovelock spaces} of order $\bar{k}$. These are a subset of {\it Einstein spaces}. The full $D$ dimensional solution (\ref{ansatz}) is then constructed by solving (\ref{genwheelereqn}) for $U(z)$ which determines the warping function $g(z)=-z^2U(z)$. 

It is worth noting that the same conditions on the base metric were obtained in \cite{Ray:2015ava} for solutions of the form of a warped product of a base space and a two-dimensional transverse space. The corresponding solutions were shown to be a by-product of Birkhoff's theorem in Lovelock theories with generic coupling constants and represent static black holes with generic horizon geometries satisfying the above conditions.

From the point of view of the $N$-dimensional brane world, the higher curvature contributions look like
cosmological constant stress-energy. That is,  on the brane one has that the Einstein tensor
is propotional to the metric,
\begin{align}\label{effective}
\hat{ G}^\alpha _{\ \ \beta} = \sum\limits_{n\neq 1}^{\bar{k}} c_n \hat{\mathcal{G}}^{(n)\alpha}_{\ \ \ \beta}=
\Lambda_{eff} \delta^{\alpha}_{\beta}.
\end{align}
where the sign of the effective cosmological constant depends on the coefficients $\alpha_n$ and the 
Lovelock couplings $c_n$.

The conditions (\ref{tensconstraint}) can also be formulated in the following alternative way.  It follows, in particular from (\ref{tensconstraint}) with $n=1$ that the base metric is an Einstein metric.  For such a metric, the Weyl tensor is given by $\hat{C}_{\mu\nu}^{\ \ \,\lambda\rho}=\hat{R}_{\mu\nu}^{\ \ \,\lambda\rho}-\alpha_1\delta_{\mu\nu}^{\lambda\rho}$, where $\alpha_1$ is the constant appearing in (\ref{tensconstraint}). Assuming that this $n=1$ constraint is satisfied, we can restate the conditions (\ref{tensconstraint}) for $n\geqslant2$ as
\begin{align}
\dfrac{1}{2^{n+1}}\delta_{\beta\nu_1\lambda_1\cdots\nu_n\lambda_n}^{\alpha\mu_1\rho_1\cdots\mu_n\rho_n}\hat{C}_{\mu_1\rho_1}^{\ \ \ \ \nu_1\lambda_1}\cdots\hat{C}_{\mu_n\rho_n}^{\ \ \ \ \nu_n\lambda_n}=\dfrac{1}{2}\dfrac{(D-2)!}{(D-2n-2)!}\lambda_n\delta_{\beta}^{\alpha}.
\end{align}
where the coefficients $\lambda_n$ are related to the coefficients in (\ref{tensconstraint}) according to
\begin{align}
\lambda_n=\sum\limits_{k=0}^{n}\binom{n}{k}(-\alpha_1)^k\alpha_{n-k}
\end{align}

The constraint (\ref{tensconstraint}) that the base metric be a Lovelock space is a strong constraint.  Note, for example, that Lovelock spaces are necessarily Einstein spaces, but the converse is certainly not true.  In particular, Schwarzschild-AdS is an Einstein space, but does not satisfy the higher order criteria in (\ref{tensconstraint}). Spaces of constant curvature satisfy the Lovelock criteria and the construction (\ref{ansatz}) in this case reproduces a higher dimensional constant curvature vacuum of the Lovelock theory (\ref{lovelock}).
Lovelock spaces also include universal spacetimes \cite{Coley:2008th,Bleecker:1979} for which all conserved symmetric rank-2 tensors contructed from the metric, the Riemann tensor and its covariant derivatives are multiples of the metric. Universal spacetimes are thus vacuum solutions to generic gravity theories with Lagrangians constructed out of the metric, the Riemann tensor and its covariant derivatives of arbitrary order and hence are also solutions to quantum-corrected theories.
Additional examples of Lovelock spaces include products of equidimensional constant curvature spaces and a certain class of homogeneous spaces \cite{Ohashi:2015xaa}.  It would be interesting to find further examples of Lovelock spaces.  



We note that special care must be taken in carrying out the analysis in odd dimensions when the order of the theory is maximal, {\it i.e.} when $D=2\bar{k}+1$, since as explained in the introduction the $\bar{k}$th order intrinsic Lovelock tensor vanishes identically on $2\bar{k}$-dimensional slices. In this case, the sum in the equation (\ref{redfeqns2}) extends only upto $n=\bar{k}-1$. Hence, following the same steps as before we conclude that there must exist a unique set of $\bar{k}$ constants $\beta_n$ for $n=0,\ldots,\bar{k}-1$, not all zero, such that
\begin{align}
\sum\limits_{n=0}^{\bar{k}-1}\beta_nl\dfrac{d\ }{dz}\left(({z\over l})^{2(\bar{k}-n)}A_n(U)\right)=0
\end{align} 
which can then be integrated to obtain
\begin{align}
\sum\limits_{n=0}^{\bar{k}-1}\beta_n\dfrac{l^{2n}}{z^{2n}}A_n(U)=C\dfrac{l^{2\bar{k}}}{z^{2\bar{k}}}
\end{align}
where $C$ is a integration constant and can be renamed as $-a_{\bar{k}}\beta_{\bar{k}}$ in order to rewrite the above equation as
\begin{align}
\sum\limits_{n=0}^{\bar{k}}\beta_n\dfrac{l^{2n}}{z^{2n}}A_n(U)=0.
\end{align}
Finally comparing with (\ref{genwheelereqn}) we conclude $\beta_n=\alpha_n$ for all $n$. On the other hand, there are now $\bar{k}-1$ tensorial conditions of the form (\ref{tensconstraint}) and one scalar constraint of the form (\ref{sclconstraint}) i.e., even dimensional base manifolds in addition to being a Lovelock space of order $\bar{k}-1$ must have a constant Euler invariant of order $\bar{k}$, as was shown in \cite{Barcelo:2002wz} for Gauss-Bonnet gravity in $D=5$.

\section{Non-generic classes of solutions}\label{polynomialsection}

Additional branches of solutions, with less restrictive conditions on the base metric can arise for certain non-generic values of the Lovelock coupling constants.  We will see that in these branches of solutions, there are fewer linearly independent relations amongst the functions on the base space appearing in (\ref{redfeqns1}) and (\ref{redfeqns2}), implying that there are correspondingly more relations amongst the functions of the $z$-coordinate.
To find these solutions, we first consider an expansion of $U(z)$ around $z=\infty$, which will have the form
\begin{align}\label{expnu}
U=-U_0-\dfrac{l^2}{z^2}U_1-\dfrac{l^4}{z^4}U_2-\dfrac{l^6}{z^6}U_3-\cdots
\end{align}
where the $U_k$ are constants and minus signs are included for convenience.  We consider the case that this expansion truncates so that the solution is given by a finite order polynomial in $1/z^2$.  In the appendix, we show that any such polynomial solution must, in fact, truncate after the first order term\footnote{We expect, though have not yet been able to prove, that solutions such that the rank of the Wronskian matrix for the set $\{A_n(U)/z^{2n}\}$ is less than $\bar{k}$ necessarily have this form.}, so that
\be\label{polynomial}
U(z) = -U_0 + {C\over z^2}
\ee
Note that the warping function is then given by
\be\label{polyg}
g(z) = U_0 z^2 - C
\ee
which is the same form that arises when adding an extra dimension in the Einstein plus cosmological
constant case discussed in the introduction (see  equation (\ref{simpleg})). The case
$U_0 >0$ corresponds to $z$ being a spacelike coordinate. When the constant $C>0$, transforming to a geodesic
coordinate gives for the metric
\be\label{simplemetric}
ds^2 = dy^2 + { |C|\over l^2 U_0 }\cosh^2 (\sqrt{U_0 } y) \hat{g}_{\alpha\beta}  dx^\alpha dx^\beta 
\ee
When $C<0$ the function $\cosh  (\sqrt{U_0 } y) $ is replaced by $\sinh (\sqrt{U_0 } y)$, and when $C=0$ one has instead the function
$e^{ \sqrt{U_0 } y}$.

Turning to the analysis of the brane metric for the special cases, 
 we now expand the functions $A_n(U)$ around $U=-U_0$ and make use of (\ref{highern}), we obtain in this case
\be
A_n(U) = \sum_{m=n}^{\bar k} \binom{m}{n} \left({C\over z^2}\right)^{m-n}A_m(-U_0)
\ee
Making use of this expression, the equations of motion (\ref{redfeqns1}) and (\ref{redfeqns2}) reduce respectively to
\begin{align}
&\sum_{m=0}^{\bar k}{l^{2m}\over z^{2m}}\left(A_m(-U_0)\label{1steqn}
\sum_{n=0}^m(D-2n-1)!\hat{\cal{L}}^{(n)}\binom{m}{n} \left({C\over l^2}\right)^{m-n}\right)=0\\
&\sum_{m=0}^{\bar k}{(D-2m-1)l^{2m}\over z^{2m}}\left(A_m(-U_0)
\sum_{n=0}^m(D-2n-2)!\hat{\cal{G}}^{(n)\alpha}{}_\beta\binom{m}{n} \left({C\over l^2}\right)^{m-n}\right)=0\label{2ndeqn}
\end{align}
In order for these equations to be satisfied, the coefficients of each power of $z$ must vanish independently.  From (\ref{2ndeqn}) we then arrive at the $\bar k+1$ conditions
\be\label{conditions}
A_m(-U_0)
\sum_{n=0}^m(D-2n-2)!\hat{\cal{G}}^{(n)\alpha}{}_\beta\binom{m}{n} \left({C\over l^2}\right)^{m-n}=0,\qquad \forall \ m=0,1,\ldots,\bar{k}
\ee
The corresponding conditions arising from (\ref{1steqn}) are proportional to the traces of these and will therefore follow from the conditions (\ref{conditions}).   In $D=2\bar{k}+1$ the $\bar{k}$th condition  is replaced by
\begin{align}\label{conditiontwo}
A_{\bar{k}}(-U_0)\sum\limits_{n=0}^{\bar{k}}\binom{\bar{k}}{n}(2\bar{k}-2n)!\hat{\mathcal{L}}^{(n)}\left({C\over l^2}\right)^{\bar{k}-n}=0
\end{align}

Each of the conditions (\ref{conditions}) can be satisfied in two different ways, by the vanishing of either the prefactor $A_m(-U_0)$ or of the sum that it multiplies.  Let us focus first on the condition for $m=0$.  Recalling from (\ref{basics}) that $\hat{\cal{G}}^{(0)\alpha}{}_\beta $ is a constant tensor, we see that in this case $A_0(-U_0)=0$ is required.  Looking back at the definition (\ref{anu}), one finds that 
\be 
A_0(\lambda) = \sum_{n=0}^{\bar k} a_n\lambda^n
\ee
which coincides with the polynomial $f(\lambda)$ in (\ref{constant}).  Therefore, the $m=0$ condition is that $\lambda=-U_0$ must be one of the allowed vacuum curvatures for the Lovelock theory (\ref{lovelock}).  Recall also from (\ref{highern}) that $A_m(U)$ with $m>0$ is proportional to the $m$th derivative of $A_0(U)$. It then follows that if  $\lambda=-U_0$ is a $p$th order root of $A_0(\lambda)$, that the prefactors $A_m(-U_0)$ of the first $p$ conditions (\ref{conditions}) will vanish, leaving $\bar k+1-p$ remaining conditions to be satisfied.  Although it is possible that further higher order derivatives of $A_0$ may also vanish at $U_0$, and we will give an example of this below, generally these remaining conditions must be satisfied by having the sum over components of the Lovelock tensors vanish, which give constraints on the base metric.

The weakest constraints on the base manifold arise in the case of a unique vacuum Lovelock theory, with couplings given by (\ref{unique}).  Taking $U_0=-\lambda_0$ in this case gives $A_m(-U_0)=0$ for $m=0,\dots,\bar k-1$.  The sum in (\ref{conditions}) must then vanish for $m=\bar k$, which for $D>2\bar{k}+1$ gives the constraint on the base metric
\begin{align}\label{luveqn}
\sum\limits_{n=0}^{\bar{k}}(D-2n-2)!\binom{\bar{k}}{n}\left({C\over l^2}\right)^{\bar{k}-n}\, \hat{\mathcal{G}}^{(n)\alpha}_{\ \ \ \beta}=0
\end{align}
This constraint has the form of the Lovelock equations of motion with a particular set of coupling constants related to the constant $C$ in the function $U(z)$ in (\ref{polynomial})\footnote{In $D=2\bar{k}+1$, the $\bar k$th constraint is instead that the base metric must satisfy
\begin{align}\label{conditionthree}
\sum\limits_{n=0}^{\bar{k}}(2\bar{k}-2n)!\binom{\bar{k}}{n}\left({C\over l^2}\right)^{\bar{k}-n}\,\hat{\mathcal{L}}^{(n)}=0.
\end{align}}.  Comparing with (\ref{unique}) one sees that the constants are in fact those of a unique vacuum theory with $\lambda_0=-C/l^2$.  Thus, if we start with a unique vacuum Lovelock theory in $D=N+1$ dimensions, we will have solutions of the form (\ref{ansatz}) with $U(z)$ as in (\ref{polynomial}) if the $N$ dimensional base metric is itself a solution to a unique vacuum Lovelock theory, with an arbitrary constant curvature\footnote{This reproduces a result originally found in \cite{Kastor:2006vw,Giribet:2006ec}.}.    Solutions to unique vacuum theories were studied in \cite{Crisostomo:2000bb} and include black hole solutions.  This mechanism then allows these to be extended to black strings, similar to the construction of AdS black strings in \cite{Chamblin:1999by}.  In terms of our general considerations in section (\ref{general}), in this case there is only a single linear relation, given by equation (\ref{luveqn}), between the functions on the base space in (\ref{redfeqns1}) and (\ref{redfeqns2}), while the functions of $z$ have $\bar k$ linearly independent relations between them and are all, in fact, constant.

The strongest constraints come if $A_m(-U_0)=0$ only for $m=0$.  It then follows from the combination of conditions (\ref{conditions}) with $m=1,\dots,\bar k$ that in dimensions $D > 2\bar{k}+1$ (or $D=2\bar{k}+1$) all the intrinsic Lovelock tensors up to order $\bar{k}$ (or $\bar{k}-1$) must be of the form
\begin{align}\label{conditionfour}
\hat{\mathcal{G}}^{(n)\alpha}_{\ \ \ \beta}=-\dfrac{1}{2}\dfrac{(D-2)!}{(D-2n-2)!}\left({-C\over l^2}\right)^n\delta^{\alpha}_{\beta}
\end{align}
This is a special case of the solutions in section (\ref{lovelockspaces}) with $\alpha_n=(-C/l^2)^n$ and is satisfied by constant curvature base spaces, which trivially are {\it Lovelock spaces} of maximal order.  In the next section, we illustrate the various possibilities for polynomial solutions in third order Lovelock theory.  This will illustrate how when more of the $A_n(-U_0)$ vanish, there will be correspondingly fewer conditions on the Lovelock tensors of the base manifold.


\section{Non-generic solutions in 3rd order Lovelock theory}\label{example}

We now illustrate the different possibilities for polynomial solutions with the example of third order Lovelock gravity.  This case captures the essential features of the analysis for any higher order.  Let $a_0,a_1,a_2$ and $a_3$ be the rescaled coupling constants (\ref{rescaled}) for the theory (\ref{lovelock}).   Depending on the ranges of these couplings, the theory can have either one or three constant curvature vacua, corresponding to real-valued solutions of (\ref{constant}).  In the case that three vacua exist, there is also the possibility of multiple roots occurring for special values of the couplings.  The relevant functions $A_m(U)$ for the 3rd order theory are
\begin{align}
A_0(U) & = a_0+a_1U +a_2U^2 +a_3U^3\\
A_1(U) &= a_1+2a_2U+3a_3U^2\\
A_2(U)&= a_2+3a_3U\\
A_3(U) &= a_3
\end{align}
Let us primarilly consider the case $D\ge 8$.  The results of the previous section then tell us that the conditions (\ref{conditions}) must be satisfied for $m=0,1,2,3$.  Following the discussion in the previous section, we know that the implications of the $m=0$ and $m=3$ constraints will not depend on the precise values of the coupling constants.
The $m=0$ condition requires that $-U_0$ be a real root of $A_0(-U_0)=0$, which coincides with the condition (\ref{constant}) for $\lambda=-U_0$ to be the curvature of a constant curvature vacuum.
The $m=3$ condition requires that the base metric be a solution for a third order unique vacuum Lovelock theory with vacuum curvature\footnote{In $D=7$ the base metric must instead satisfy the scalar constraint 
\begin{align}
720\left(C\over l^2\right)^3\hat{\mathcal{L}}^{(0)}+72\left(C\over l^2\right)^2\hat{\mathcal{L}}^{(1)}+6\left(C\over l^2\right)\hat{\mathcal{L}}^{(2)}+\hat{\mathcal{L}}^{(3)}=0.
\end{align}
} 
$\lambda_0=-C/l^2$,
\begin{align}\label{luvcondition}
(D-2)!\left(C\over l^2\right)^3\hat{\mathcal{G}}^{(0)\alpha}{}_{\beta}&+3(D-4)!\left(C\over l^2\right)^2\hat{\mathcal{G}}^{(1)\alpha}{}_{\beta}
\\ \nonumber
&+3(D-6)!\left(C\over l^2\right)\hat{\mathcal{G}}^{(2)\alpha}{}_{\beta}+(D-8)!\hat{\mathcal{G}}^{(3)\alpha}{}_{\beta}=0
\end{align}
How the conditions (\ref{conditions}) for $m=1,2$ may be satisfied, however, do depend on the precise values of the coupling constants and fall into one of the four following cases depending on whether any or all of the constants $A_1(-U_0)$ and $A_2(-U_0)$ vanish.
\begin{enumerate}

\item If $A_1(-U_0)\neq 0$ and $A_2(-U_0)\neq 0$, then the vacuum curvature $-U_0$ is not a multiple root and all the sums in conditions (\ref{conditions}) for $m=1,2$ must also vanish. This is the most constrained case, discussed above, in which $m=1,2$ conditions in combination with those above imply that the base metric must be a Lovelock space of order $3$, with the Lovelock tensors given by (\ref{conditionfour}) in dimensions $D\geq 8$, or of order $2$ along with satisfying the scalar constraint mentioned above in dimensions $D=7$.
\item  If $A_1(-U_0)= 0$ and $A_2(-U_0)\neq 0$, then $-U_0$ is a doubly degenerate root. 
In this case, the $m=1$ condition is satisfied identically without any additional restriction on the base metric, while the $m=2$ condition implies that the corresponding sum in (\ref{conditions}) must vanish. This implies that in addition to satisfying (\ref{luvcondition}) the base metric must simultaneously be a solution to a second order unique vacuum Lovelock theory with field equations, also with vacuum curvature $-C/l^2$
\begin{align}
(D-2)!\left(C\over l^2\right)^2\,\hat{\mathcal{G}}^{(0)\alpha}{}_{\beta}+2(D-4)!\left(C\over l^2\right)\,\hat{\mathcal{G}}^{(1)\alpha}{}_{\beta}+(D-6)!\hat{\mathcal{G}}^{(2)\alpha}{}_{\beta}=0
\end{align}

\item  If both $A_1(-U_0)= 0$ and $A_2(-U_0)=0$, then $-U_0$ is a triply degenerate root. 
This will be the case if the Lovelock couplings take the values of a unique vacuum theory (\ref{unique}) with vacuum curvature $\lambda_0=-U_0$.
In this case the $m=1$ and $m=2$ conditions are identically satisfied and (\ref{luvcondition}) is the only constraint on the base manifold.

\item  If $A_1(-U_0)\neq 0$ and $A_2(-U_0)=0$, then $-U_0$ is only a single root of $A_0(U)$.  However, the Lovelock coefficients are such that the $2$nd derivative of $A_0(U)$ also vanishes at that point. In this case the $m=2$ condition is identically satisfied, while the $m=1$ condition implies that the base manifold must also be an Einstein space satisfying
\begin{align}
(D-2)!\left(C\over l^2\right)\,\hat{\mathcal{G}}^{(0)\alpha}_{\ \ \ \beta}+(D-4)!\hat{\mathcal{G}}^{(1)\alpha}_{\ \ \ \beta}=0
\end{align}
\end{enumerate}
These cases for third order Lovelock gravity are indicative of the different cases occurring in higher order theories; the possibility of multiple roots and the possible vanishing of still higher order derivatives of the function $A_0(U)$ at $U=-U_0$.

\section{Conclusions}\label{conclusions}

We have shown that $D=N+1$ dimensional solutions to Lovelock gravity theories of the form (\ref{ansatz}) may be constructed starting from different classes of $N$-dimensional base metrics. 
 For Lovelock theories with generic values of the coupling constants, the Lovelock tensors of the base metric must satisfy the constraints (\ref{tensconstraint}), which imply that they are all constant multiples of the metric.  This condition generalizes the notion of an Einstein space, and we take this to
be the defining property of a `Lovelock space'.  

For special values of the Lovelock couplings, the constraints on the base manifold are weaker.  The most extreme, and most interesting, example of this arises in unique vacuum theories.  In this case, the Lovelock tensors need satisfy only a single constraint (\ref{luveqn}), which implies that the base metric is itself a solution of an $N$-dimensional unique vacuum Lovelock theory, whose vacuum curvature is independent of that of the $D$-dimensional theory, reproducing an earlier result \cite{Kastor:2006vw,Giribet:2006ec}.   In this case, since the $N$-dimensional theory admits black hole solutions, $D=N+1$ \cite{Giribet:2006ec}, the construction (\ref{ansatz}) yields black strings.  There are also intermediate cases between the generic Lovelock theory and the unique vacuum theory, such that the Lovelock tensors also satisfy fewer constraints than (\ref{tensconstraint}).  The possibilities for such special cases have been illustrated in detail in third order Lovelock gravity.

Our emphasis here has been on establishing these constraints on the base metric.  It would also be interesting to discover additional examples of spaces satisfying these constraints, all of which are generalizations, in one way or another, of the notion of an Einstein space.

{\bf Acknowledgements}

The work of SR is supported by FONDECYT grant 1150907.

\section*{Appendix}

\subsection*{Asymptotic expansion}
Here we shall analyze the asymptotic form of the function $U(z)$ solving equation (\ref{genwheelereqn}) for generic coupling constants $a_k$. This is a polynomial equation in $U$ of order $\bar{k}$ with coefficients depending on $z^2$. So, the general solution is given in the form of radicals which can be expanded about $1/z^2$ as
\begin{align}\label{expnu2}
U=-U_0-\dfrac{l^2}{z^2}U_1-\dfrac{l^4}{z^4}U_2-\dfrac{l^6}{z^6}U_3-\cdots
\end{align}
where the coefficents now depend on $\alpha_n$ and $a_n$. These asymptotic coefficients can be found by solving (\ref{genwheelereqn}) term by term. The first coefficient $U_0$ characterizes the (inverse squared) curvature radius at $z \to \infty$ and is given by one of the non-degenerate real roots of the polynomial $A_0(-U)$. The next few coefficients are given as
\begin{align}
U_1=&\alpha_1\\
U_2=&(\alpha_2-\alpha_1^2)\dfrac{A_2(-U_0)}{A_1(-U_0)}\\
U_3=&(\alpha_3-3\alpha_2\alpha_1+2\alpha_1^3)\dfrac{A_3(-U_0)}{A_1(-U_0)}\\
U_4=&(\alpha_4-4\alpha_3\alpha_1+6\alpha_2\alpha_1^2-3\alpha_1^4)\dfrac{A_4(-U_0)}{A_1(-U_0)}\\
&+(\alpha_2-\alpha_1^2)^2\dfrac{A_2(-U_0)}{A_1(-U_0)^3}\left(A_2(-U_0)^2-3A_3(-U_0)A_1(-U_0)\right)
\end{align}
and so on.
Suppose there are a finite number of terms in the expansion (\ref{expnu}). If $-U_m$ is the coefficient of the highest order term in (\ref{expnu}) with $m>1$, then expanding the left hand side of (\ref{genwheelereqn}), the coefficient of the highest order term  is found to be $a_{\bar{k}}(-U_m/z^{2m})^{\bar{k}}$. Equating this to zero we infer $U_m=0$. In other words, the only possible polynomial form of the solution to (\ref{genwheelereqn}) is 
\begin{align}
U=-U_0-\dfrac{l^2}{z^2}U_1
\end{align}

%
%

\end{document}